# Brain-to-Speech: Prosody Feature Engineering and Transformer-Based Reconstruction

Mohammed Salah Al-Radhi, Géza Németh, Andon Tchechmedjiev, and Binbin Xu

**Abstract** This chapter presents a novel approach to brain-to-speech (BTS) synthesis from intracranial electroencephalography (iEEG) data, emphasizing prosody-aware feature engineering and advanced transformer-based models for high-fidelity speech reconstruction. Driven by the increasing interest in decoding speech directly from brain activity, this work integrates neuroscience, artificial intelligence, and signal processing to generate accurate and natural speech. We introduce a novel pipeline for extracting key prosodic features directly from complex brain iEEG signals, including intonation, pitch, and rhythm. To effectively utilize these crucial features for natural-sounding speech, we employ advanced deep learning models. Furthermore, this chapter introduces a novel transformer encoder architecture specifically designed for brain-to-speech tasks. Unlike conventional models, our architecture integrates the extracted prosodic features to significantly enhance speech reconstruction, resulting in generated speech with improved intelligibility and expressiveness. A detailed evaluation demonstrates superior performance over established baseline methods, such as traditional Griffin-Lim and CNN-based reconstruction, across both quantitative and perceptual metrics. By demonstrating these advancements in feature extraction and transformer-based learning, this chapter contributes to the growing field of AI-driven neuroprosthetics, paving the way for assistive technologies that restore communication for individuals with speech impairments. Finally, we discuss promising future research directions, including the integration of diffusion models and real-time inference systems.

**Keywords** Brain AI · Neural signals · iEEG · Prosody feature · Transformer model

M. S. Al-Radhi (✉) · G. Németh
Department of Telecommunications and Artificial Intelligence, Budapest University of Technology and Economics, Budapest, Hungary
e-mail: malradhi@tmit.bme.hu; nemeth@tmit.bme.hu

A. Tchechmedjiev · B. Xu
EuroMov Digital Health in Motion, University of Montpellier, IMT Mines Alès, Montpellier, France
e-mail: andon.tchechmedjiev@mines-ales.fr; binbin.xu@mines-ales.fr

## 1 Introduction

Speech is a fundamental means of human communication, and its loss due to neurological disorders can severely impact quality of life. Brain-to-speech (BTS) technology, which aims to synthesize speech directly from neural activity, has emerged as a promising solution for individuals with speech impairments [1–4]. Among various neural recording techniques [5], intracranial electroencephalography (iEEG) provides high spatial and temporal resolution, making it an ideal candidate for reconstructing speech from brain signals [6]. However, despite recent advances in deep learning and neural decoding [7–10], achieving natural and intelligible speech synthesis remains a significant challenge.

A major limitation in current BTS research is the inadequate modeling of prosody [11, 12], which encompasses rhythm, stress, and intonation (key elements that contribute to speech naturalness and intelligibility). Traditional speech synthesis techniques primarily focus on spectral envelope reconstruction [13], often neglecting these essential prosodic attributes, resulting in robotic and monotonous speech. Furthermore, deep learning models struggle with the inherent variability in neural signals [14, 15], limiting their ability to generalize across subjects and speech conditions [16, 17].

Another critical challenge is the accurate reconstruction of phase information, which is essential for high-fidelity speech synthesis. Traditional vocoding techniques, such as the Griffin-Lim algorithm, introduce phase inconsistencies, leading to perceptual distortions [18, 19]. Although neural vocoders have demonstrated promising improvements, their adaptation to iEEG-based speech synthesis remains underexplored.

To address these challenges, this chapter introduces a novel BTS framework that integrates prosody-aware feature extraction with a transformer-based speech reconstruction model. The proposed approach employs a wavelet-based feature extraction pipeline to capture both spectral and prosodic characteristics of iEEG signals, enhancing their representation for speech decoding. In addition, a specialized transformer encoder architecture is designed to control the extracted prosodic features, significantly improving speech synthesis quality by generating more natural and meaningful speech. To further enhance phase reconstruction, an iterative harmonic phase correction mechanism is introduced, ensuring greater harmonic consistency and minimizing perceptual distortions. The effectiveness of this framework is evaluated against state-of-the-art iEEG-to-speech synthesis models, demonstrating superior performance in terms of intelligibility and naturalness.

Beyond the immediate contributions to BTS synthesis, this work provides insights into the broader applications of brain-computer interface (BCI) research, particularly in the development of neural speech prostheses. By advancing prosody-aware feature engineering and transformer-based reconstruction methods, this chapter contributes to the growing field of AI-driven neuroprosthetics, paving the way for future innovations in assistive communication technologies. Moreover, potential directions for future research include the integration of diffusion-based generative models and real-time inference strategies, which could further enhance the feasibility and practicality of BTS systems. Additionally, this chapter aligns with the objectives of the AI, Data, and Robotics Partnership [20], contributing to advancements in AI-driven speech neuroprostheses and data-driven methodologies for brain-computer interfaces.

The remainder of this chapter is organized as follows: Sect. 2 reviews related work in brain-to-speech synthesis, focusing on feature extraction, deep learning models, and vocoding techniques. Section 3 presents the proposed methodology, detailing the prosody feature engineering pipeline and the transformer-based reconstruction model. Section 4 describes the experimental setup, including datasets, evaluation metrics, and baseline comparisons. Section 5 discusses results and provides insights into model performance. Finally, Sect. 6 outlines future research directions and concludes the chapter.

## 2   Related Work

The field of brain-to-speech (BTS) synthesis has evolved significantly over the past decade, supported by advances in neural signal processing, deep learning, and speech synthesis technologies. Early studies primarily focused on linear mappings between neural activity and acoustic features such as formants or spectral envelopes [21], which provided foundational insights but lacked the capacity to model the complex temporal and spectral structures inherent in natural speech. With the rise of deep learning, more sophisticated models have been developed to decode speech directly from neural signals, improving intelligibility and expressiveness.

### 2.1   Existing Deep Learning Models for Brain-to-Speech

Several deep learning architectures have been explored to map neural recordings to speech features. Convolutional neural networks (CNNs) [22] have been widely used due to their ability to capture local spatial and spectral structures in iEEG data. CNNs have shown improved accuracy in reconstructing speech by focusing on features like high-gamma power and band-specific energy distributions [23].

Recurrent neural networks (RNNs), particularly long short-term memory (LSTM) networks [24], have also been employed to model temporal dependencies in neural signals, which are essential for prosody and rhythm. However, RNNs often struggle with long-range dependencies and generalization across speakers and sessions. Sequence-to-sequence (Seq2Seq) models [25], enhanced with attention mechanisms, have demonstrated more robust performance by learning both local and global temporal relationships. Despite their benefits, many Seq2Seq approaches still lack explicit prosody modeling, limiting the expressiveness of the synthesized speech.

## 2.2  Feature Extraction Techniques in iEEG-to-Speech Studies

Accurate representation of neural features is critical in BTS systems. Prior studies have commonly used empirical features such as high-gamma activity, band-limited power, and phase-amplitude coupling (PAC) to capture speech-relevant neural dynamics [23]. High-gamma activity (70–170 Hz) has been associated with articulatory processing, while slower oscillations in the theta and beta ranges have been linked to prosodic elements and motor planning [9].

To improve temporal resolution and multi-scale analysis, wavelet-based methods have been introduced. Discrete Wavelet Transform (DWT), for example, allows decomposition of iEEG signals into time-frequency representations that can capture both transient articulatory and sustained prosodic components [26]. Furthermore, cross-frequency coupling (CFC) metrics, particularly PAC, have been leveraged to understand hierarchical interactions between frequency bands in speech production [27].

## 2.3  Prior Work on Prosody Representation in BTS Systems

Prosody, which includes intonation, stress, and rhythm, is crucial for producing natural-sounding speech, yet it has historically been underrepresented in BTS literature. Many earlier models focused solely on spectral envelope reconstruction, often yielding robotic outputs lacking emotional tone or natural flow.

Recent research has begun to address this by incorporating prosodic cues directly into neural decoding pipelines. Some studies have attempted to predict fundamental frequency (F0) from neural recordings using supervised learning techniques [28], while others have explored additional prosodic features such as energy, duration, shimmer (amplitude perturbation), and phase variability [29]. However, consistent integration of these features into BTS models remains limited, and more comprehensive approaches to prosody-aware decoding are still needed.

## 2.4  Neural Vocoding Methods Applied in BTS Research

Generating high-fidelity audio from predicted acoustic features is a core challenge in BTS systems. Traditional vocoders like the Griffin-Lim algorithm [30] have been used to synthesize waveforms from spectrograms, but they often introduce artifacts due to inaccurate phase estimation.

To overcome these limitations, researchers have adapted neural vocoders such as WaveGlow [31], BigVGAN [32], and AutoVocoder [33], which use deep generative models to synthesize more natural-sounding speech. These models have proven effective in text-to-speech (TTS) and speech enhancement applications [34, 35], though their direct application to iEEG-based BTS systems remains limited. Further exploration is needed to tailor neural vocoders for the unique properties of brain-derived acoustic features.

## 3  Proposed Framework

Decoding speech directly from intracranial EEG (iEEG) recordings presents a significant challenge due to the complexity of neural activity and the intricate relationship between brain signals and speech production. This section introduces a novel brain-to-speech (BTS) synthesis framework that integrates prosody-aware neural encoding, transformer-based spectrogram prediction, and an iterative harmonic phase reconstruction vocoder to achieve highly intelligible and natural speech. By leveraging wavelet-based feature extraction techniques and deep learning methodologies as shown in Fig. 1, the proposed approach effectively captures neural correlates of speech production; models prosodic attributes such as pitch, rhythm, and stress; and ensures phase-consistent synthesis for high-fidelity neural speech reconstruction.

### 3.1  Prosody Embedding

The first component of the proposed model extracts multi-modal features from iEEG signals while embedding prosodic information to capture the temporal, spectral, and neurophysiological dynamics of speech production. This step bridges neural activity and speech synthesis by ensuring that the extracted features are both discriminative and representative of speech-related processes.

#### 3.1.1  Wavelet-Based iEEG Representation

Neural activity associated with speech production is highly dynamic, exhibiting fluctuations across multiple time scales and frequency bands. Conventional feature extraction methods rely on fixed-band spectral features, which often fail to capture the rich temporal and spectral variations necessary for accurate speech synthesis. To overcome this limitation, the proposed framework employs Discrete Wavelet Transform (DWT), a powerful signal processing technique that decomposes iEEG signals into multiple resolution levels, enabling multi-scale analysis of neural dynamics.

The Daubechies-4 (db4) wavelet is selected due to its superior ability to analyze non-stationary and transient signals, which are characteristic of iEEG recordings. Given an iEEG signal $x(t)$, the wavelet decomposition can be expressed as follows:

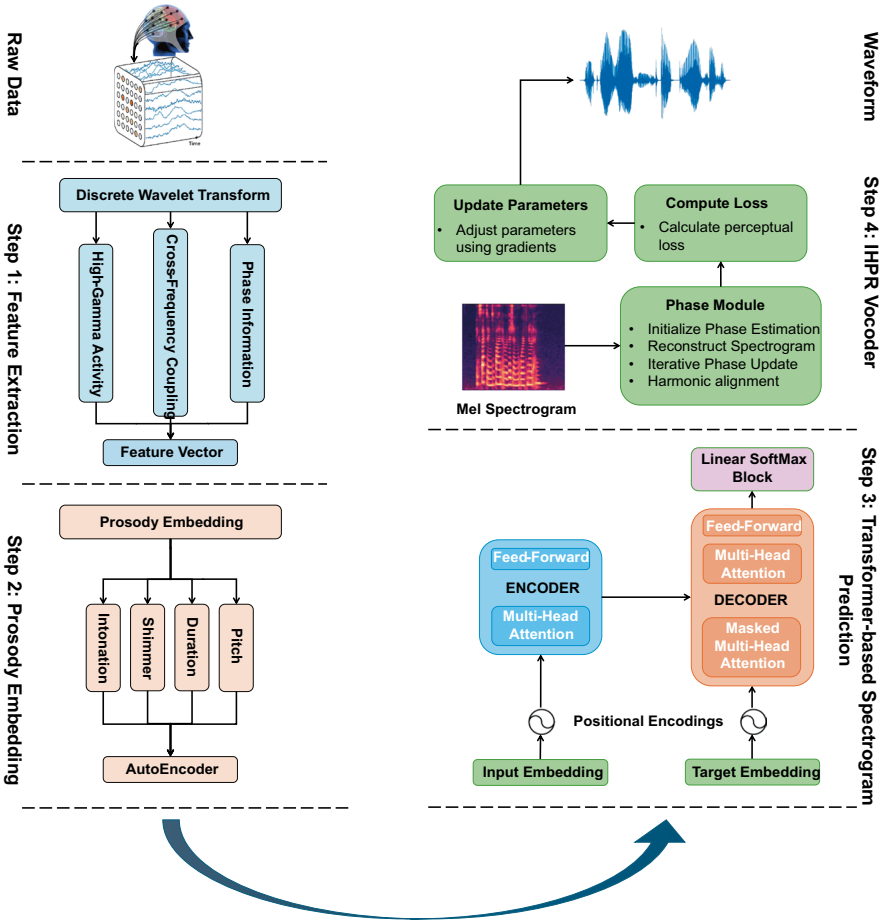

**Fig. 1** Schematic diagram of the proposed architecture

$$x(t) = \sum_j \sum_n c_j(n) \psi_{j,n}(t) \tag{1}$$

where $c_j(n)$ are the wavelet coefficients at scale $j$ and position $n$, $\psi_{j,n}(t)$ represents the wavelet basis function. The coefficients capture both high-frequency articulatory dynamics and low-frequency prosodic modulations.

The energy distribution across wavelet scales provides insights into speech-relevant neural oscillations:

$$E_j = \sum_n |c_j(n)|^2 \tag{2}$$

where $E_j$ represents the total energy at scale $j$, enabling identification of dominant neural patterns in different frequency bands. By leveraging this multi-scale wavelet representation, we extract speech-relevant neural patterns from distinct frequency bands:

- **High-gamma (70–170 Hz)**: Captures phoneme articulation and fine-grained speech motor control
- **Beta (15–30 Hz)**: Reflects motor planning and coordination
- **Theta (4–8 Hz)**: Encodes syllabic timing and prosodic modulation

Each wavelet coefficient set $c_j(n)$ is then mapped into a feature space, where prosodic and articulatory information are encoded separately. To ensure that prosodic features (intonation, rhythm, stress) are effectively captured, we introduce a prosody embedding layer, which transforms the extracted wavelet features into a latent representation suitable for speech synthesis. This is performed as follows:

$$P = f_{\text{prosody}}(c_\theta, c_\beta) \tag{3}$$

where $c_\theta$ and $c_\beta$ are wavelet coefficients from theta and beta bands, encoding prosodic variations, and $f_{\text{prosody}}$ is a non-linear mapping function (e.g., a neural network) that embeds the extracted rhythm and stress patterns into a lower-dimensional space. The resulting vector $P$ serves as an input to the speech synthesis model, ensuring intelligible and natural speech output.

This multi-scale feature extraction and embedding process enhances speech reconstruction by preserving both fine-grained articulatory features and global prosodic dynamics, leading to more natural and nuanced speech synthesis.

### 3.1.2 Cross-Frequency Coupling (CFC) Analysis

The process of speech production is controlled by hierarchical neural oscillations, where low-frequency cortical rhythms modulate high-frequency neural activity. This cross-frequency interaction plays a crucial role in encoding syllabic structure, prosody, and phonemic transitions. To capture these relationships, the proposed framework employs Phase-Amplitude Coupling (PAC), a well-established measure that quantifies the degree to which the phase of low-frequency oscillations modulates the amplitude of high-frequency activity:

$$\text{PAC}(t) = \left| E\left[ A_\Upsilon(t) e^{j\phi_\theta(t)} \right] \right| \tag{4}$$

where $A_\Upsilon(t)$ is the amplitude envelope of high-gamma activity, $\phi_\theta(t)$ is the phase of theta-band oscillations extracted using the Hilbert transform, and $E[\cdot]$ denotes the expectation operator.

PAC features are extracted to enhance temporal alignment between neural oscillatory patterns and speech events, ensuring that the generated speech preserves the timing and rhythmic structure present in natural human communication. By incorporating PAC analysis, the framework is able to synchronize prosodic patterns with the temporal evolution of neural activity, leading to more fluent and natural speech synthesis.

### 3.1.3 Prosodic Feature Extraction and Normalization

Natural speech is inherently prosodic, meaning that variations in pitch, loudness, and duration play a critical role in conveying meaning and emotion. Many existing brain-to-speech systems neglect these aspects, resulting in monotonic and robotic speech synthesis. To address this limitation, the proposed framework explicitly extracts prosody-related features from iEEG signals, including:

- **Fundamental frequency (F0)**: Estimated using the Harvest algorithm [36], capturing intonation and pitch contours
- **Energy (loudness)**: Computed as the root mean square (RMS) of the neural signal, modeling speech emphasis and stress patterns
- **Shimmer**: A measure of frame-to-frame amplitude variability, used to enhance the natural dynamics of speech synthesis
- **Duration**: Encodes the temporal structure of phonemes and syllables, ensuring realistic speech timing
- **Phase variability**: The standard deviation of instantaneous phase fluctuations

These features are computed using a 50 ms analysis window with a 10 ms frame-shift, ensuring fine-grained temporal alignment with the iEEG signals. To maintain consistency across subjects and recording conditions, all extracted features undergo *z*-score normalization:

$$\hat{x} = \frac{x - \mu}{\sigma} \tag{5}$$

where $\mu$ and $\sigma$ represent the mean and standard deviation of each feature. This normalization ensures that prosodic attributes are consistently represented, improving the model's ability to generate expressive, speaker-independent speech reconstructions.

## 3.2 Transformer-Based Spectrogram Reconstruction

This component predicts Mel spectrograms from encoded neural features through two stages: (1) dimensionality reduction using an autoencoder and (2) spectrogram prediction using a transformer model to capture long-range dependencies and temporal dynamics.

### 3.2.1 Autoencoder-Based Latent Feature Encoding

Neural data is high dimensional and noisy, making direct mapping to speech representations challenging. To overcome this, the framework employs an autoencoder-based feature compression module, which learns a compact, information-rich latent space from iEEG features.

The encoder component of the autoencoder consists of fully connected layers with ReLU activations, transforming the raw iEEG features into a lower-dimensional latent vector while preserving essential speech-related information. The decoder reconstructs the input by minimizing a Mean Squared Error (MSE) loss, ensuring that the most essential neural features are retained for spectrogram prediction. This feature compression step improves the model's ability to generalize across different speech patterns while reducing computational complexity. After training using the Adam optimizer with a 0.001 learning rate, the encoder is used to generate latent representations that serve as the input to the transformer-based spectrogram predictor, allowing the model to efficiently capture complex dependencies in neural data without the risk of information loss.

### 3.2.2 Self-Attention-Based Spectrogram Prediction

Mapping neural activity to speech requires a model that can effectively capture long-range dependencies and hierarchical relationships in speech data. While RNN-based models such as LSTMs and GRUs have been explored for this task [24, 25], they suffer from limited memory capacity, making it difficult to learn dependencies over long time spans, and sequential processing constraints, which reduce efficiency and scalability. To overcome these limitations, our framework employs a Transformer-based spectrogram predictor, which leverages self-attention mechanisms to dynamically learn contextual dependencies between neural features and spectrogram frames.

Unlike RNNs, the Transformer processes entire sequences in parallel, improving efficiency and enabling long-range dependencies to be learned without memory loss. Notably, the multi-head self-attention mechanism allows the model to capture fine-grained spectral variations by attending to different temporal and spectral features simultaneously. Since Transformers lack an inherent notion of sequence order, a positional encoding function is applied to introduce temporal information:

$$\text{PE}(t, 2i) = \sin\left(\frac{t}{10{,}000^{\frac{2i}{d}}}\right) \tag{6}$$

$$\text{PE}(t, 2i+1) = \cos\left(\frac{t}{10{,}000^{\frac{2i}{d}}}\right) \tag{7}$$

where *t* represents the time step and *d* is the feature dimension.

The core of the Transformer is the multi-head self-attention mechanism, which allows the model to attend to different aspects of the input simultaneously. Given an input sequence *X*, the attention weights are computed as:

$$\text{Attention}(Q,K,V) = \text{softmax}\left(\frac{QK^T}{\sqrt{d_k}}\right)V \tag{8}$$

where *Q*, *K*, and *V* represent the query, key, and value matrices and $d_k$ is the dimension of the key vectors. Multi-head attention enables parallel attention over multiple feature subspaces, ensuring that both phonetic and prosodic cues are preserved. Each attention block is followed by a position-wise feedforward network and layer normalization, which stabilizes training and improves convergence.

The Transformer is trained using Mean Squared Error (MSE) loss between the predicted and ground-truth spectrograms, with the Adam optimizer set to a learning rate of 0.001. To prevent overfitting, dropout with a rate of 0.1 is applied during training.

The final output is a high-resolution Mel spectrogram, which serves as the input to a neural vocoder for waveform reconstruction. By integrating self-attention mechanisms, our model ensures that speech dynamics, prosody, and spectral details are accurately preserved, leading to highly intelligible and natural speech synthesis.

### 3.2.3 Iterative Harmonic Phase Reconstruction (IHPR) Vocoder

While traditional vocoders such as Griffin-Lim [30] can reconstruct speech from spectrograms, they often introduce phase inconsistencies, which lead to perceptual distortions and unnatural speech quality. To overcome this, we introduce an Iterative Harmonic Phase Reconstruction (IHPR) vocoder that enforces harmonic constraints on the phase estimation process.

The core idea of IHPR is to iteratively refine the Short-Time Fourier Transform (STFT) phase estimates to preserve the natural harmonic structure of speech. Given an initial phase estimate $\emptyset_k(t,f)$, the phase at iteration *k* + 1 is updated as:

$$\phi_{k+1}(t,f) = \text{argarg} \sum_{h=1}^{H} \cos\left(\phi(t,f_h) - \angle S_k(t,f_h)\right) \tag{9}$$

where $f_h = h \cdot f_0$ represents the *h*th harmonic frequency (with $f_0$ being the estimated fundament frequency) and $\angle \hat{S}_k(t,f_h)$ is the estimated phase angle of the STFT at harmonic $f_h$ and iteration *k*.

To further reduce phase discontinuities, we introduce a spectral correction term using a derivative-based smoothing strategy:

$$\phi_{k+1}(t,f) = \phi_k(t,f) - \lambda \sum_{h=1}^{H} \frac{\partial}{\partial f}\left[ M(t,f_h) \cdot e^{j\phi_k(t,f_h)} \right] \quad (10)$$

where $M(t,f_h)$ is the magnitude of the STFT at time $t$ and frequency $f_h$ and $\lambda$ is a smoothing coefficient that controls the strength of the correction.

To ensure convergence, a perceptual loss is used:

$$L_{\text{perceptual}} = \sum_{t,f} w(f) \left| M_{\text{target}}(t,f) - M_{\text{reconst}}(t,f) \right|^2 + \gamma \sum_{h=1}^{H} \left| \phi_k(t,f_h) - \phi_{k-1}(t,f_h) \right|^2 \quad (11)$$

where $w(f)$ is a frequency-dependent weighting function, $\gamma$ is a regularization parameter to stabilize phase evolution, and $M_{\text{target}}$ and $M_{\text{reconst}}$ are the target and reconstructed STFT magnitudes, respectively.

By integrating these phase refinement techniques, the vocoder significantly improves the perceptual quality of the synthesized speech, achieving higher harmonic-to-noise ratios and reduced spectral distortion compared to baseline models.

## 4 Experimental Design

To assess the performance of the proposed brain-to-speech (BTS) synthesis framework, an extensive experimental design was established, integrating a high-resolution intracranial EEG (iEEG) dataset, state-of-the-art deep learning methodologies, and objective and perceptual evaluation metrics. This experimental setup aimed to evaluate the framework's ability to generate natural and intelligible speech from neural activity.

### *4.1 Dataset and Preprocessing*

The study utilized a publicly available iEEG dataset[1] [13] recorded from ten participants with pharmaco-resistant epilepsy (mean age 32 years, five male, five female, and native speakers of Dutch), who are undergoing intracranial stereotactic EEG (sEEG) monitoring as part of their clinical treatment. Each participant was instructed to produce a set of isolated words and continuous speech, allowing for the collection of simultaneous neural and acoustic recordings. The iEEG signals were recorded using multi-contact depth electrodes implanted in speech-relevant cortical areas, including the superior temporal gyrus (STG), sensorimotor cortex (SMC), and inferior frontal gyrus (IFG). These regions are known to encode phonemic, articulatory, and prosodic information, making them highly suitable for brain-to-speech decoding.

---

[1] https://osf.io/nrgx6/

To preserve the fine-grained temporal structure of neural signals, the iEEG recordings were sampled at 1024 Hz, ensuring high-resolution capture of rapid speech-related neural fluctuations. Simultaneously, the speech waveforms were recorded at 16 kHz using a high-fidelity microphone, maintaining acoustic clarity for precise alignment with neural activity. Given the susceptibility of iEEG signals to various noise sources, a multi-stage preprocessing pipeline was applied to enhance data quality. First, a bandpass filter (0.5–170 Hz) was used to remove slow drifts and high-frequency artifacts. Next, notch filtering at 50 Hz and its harmonics were employed to eliminate power line interference. To standardize neural feature distributions, *z*-score normalization was applied per electrode, ensuring that feature scales remained consistent across subjects.

A critical aspect of preprocessing involves precise neural-speech alignment, as accurate time synchronization is essential for effective model training. Speech onset markers were extracted from the audio recordings using energy-based voice activity detection (VAD), while corresponding neural segments were identified through cross-correlation analysis. This alignment ensured that each neural time window corresponded precisely to the intended phonemes, allowing for robust iEEG-to-speech mapping. Also, trials containing excessive motion artifacts or electromyographic (EMG) contamination were removed, ensuring that the dataset retained only high-quality neural-speech pairs for training and evaluation.

### *4.2 Model Training and Implementation*

Following preprocessing, a multi-modal feature extraction pipeline was implemented to obtain a comprehensive representation of speech-related neural dynamics. The feature extraction process leveraged discrete wavelet transform (DWT) to decompose iEEG signals into multiple frequency bands, allowing the capture of both high-frequency articulatory patterns and low-frequency prosodic modulations. The resulting feature set incorporated high-gamma power (70–170 Hz), which has been strongly associated with phoneme articulation, as well as cross-frequency coupling (CFC) features, which quantify interactions between theta (4–8 Hz) and gamma activity, crucial for encoding rhythmic speech elements. Prosody features such as pitch (F0), intensity, shimmer, and duration were also extracted from iEEG signals, enriching the model with intonation and stress information to enhance speech naturalness.

To map neural features to a time-frequency representation of speech, a transformer-based spectrogram prediction model was employed. The first stage of this model consisted of an autoencoder-based latent feature encoding module [37], which compressed high-dimensional neural features into a compact latent space, reducing redundancy while preserving key speech-related attributes. The encoder component utilized fully connected layers with rectified linear unit (ReLU) activations, transforming the input into a low-dimensional representation, while the decoder reconstructed the original feature space with minimal information loss using mean squared error (MSE) optimization. This latent representation was then used as input to a self-attention-based transformer model, designed to capture long-range dependencies between neural activity and speech spectrogram frames. Unlike recurrent architectures such as LSTMs, which process sequences sequentially, the transformer model operates in parallel, enabling faster training and more effective feature integration.

To reconstruct the waveform from the predicted spectrograms, an Iterative Harmonic Phase Reconstruction (IHPR) vocoder was developed, addressing limitations associated with conventional phase estimation techniques. Unlike standard vocoders such as Griffin-Lim, which introduce phase artifacts and spectral distortions, the proposed IHPR vocoder enforced harmonic consistency across frequency bands using adaptive phase correction strategies.

The full model[2] was trained using the Adam optimizer with a learning rate of 0.001, employing a tenfold cross-validation protocol to ensure robust performance estimation. Training was conducted on NVIDIA A100 GPUs, utilizing accelerated deep-learning frameworks to optimize computation time and model efficiency.

### 4.3 Evaluation Metrics

The effectiveness of the proposed brain-to-speech framework was assessed using a combination of objective and perceptual evaluation metrics. Wherever possible, we present the mathematical formulations below.

#### 4.3.1 Pearson Correlation Coefficient (PC)

To measure the similarity between the predicted Mel spectrogram $\hat{Y}$ and the ground-truth spectrogram $Y$, the Pearson correlation coefficient is computed as:

$$\text{PC} = \frac{\sum_{i=1}^{N}(Y_i - \underline{Y})(\hat{Y}_i - \underline{\hat{Y}})}{\sqrt{\sum_{i=1}^{N}(Y_i - \underline{Y})^2} \cdot \sqrt{\sum_{i=1}^{N}(\hat{Y}_i - \underline{\hat{Y}})^2}} \tag{12}$$

where $N$ is the number of frames and $\underline{Y}$ and $\underline{\hat{Y}}$ are the means of the true and predicted spectrogram values, respectively. Higher PC values indicate stronger correlation and thus better intelligibility.

---

[2] The project code will be released upon acceptance.

### 4.3.2 Mel Cepstral Distortion (MCD)

The MCD is used to evaluate the spectral distance between the predicted and ground-truth Mel cepstral coefficients. It is computed as:

$$\text{MCD} = \frac{10}{\ln\ln 10} \cdot \sqrt{2\sum_{d=1}^{D}(c_d - \hat{c}_d)^2} \qquad (13)$$

where $c_d$ and $\hat{c}_d$ are the $d$th coefficients of the target and predicted Mel cepstra and $D$ is the number of cepstral dimensions. Lower MCD values indicate less spectral distortion.

### 4.3.3 Short-Time Objective Intelligibility (STOI)

STOI assesses the speech intelligibility by comparing short-time spectral features. The core idea is to compute the correlation between temporal envelopes of short-time spectral bands of the clean and synthesized speech:

$$\text{STOI} = \frac{1}{T}\sum_{t=1}^{T}\text{corr}(X_t, \hat{X}_t) \qquad (14)$$

where $X_t$ and $\hat{X}_t$ are the clean and degraded short-time spectral representations at frame $t$ and $T$ is the total number of frames. STOI returns a score between 0 and 1, where higher is better.

### 4.3.4 Harmonic-to-Noise Ratio (HNR)

HNR measures the ratio between the periodic (harmonic) and non-periodic (noise) components of the synthesized speech signal:

$$\text{HNR}(\text{dB}) = 10 \cdot \log_{10}\left(\frac{P_{\text{harmonic}}}{P_{\text{noise}}}\right) \qquad (15)$$

where $P_{\text{harmonic}}$ is the power of the harmonic signal and $P_{\text{noise}}$ is the power of the noise component. Higher HNR values suggest better phase reconstruction and more natural speech.

### 4.3.5 MOSA-Net (Perceptual Evaluation)

Beyond these standard measures, MOSA-Net [38] was used as a deep-learning-based non-intrusive speech quality model. It integrates CNNs, LSTMs, and self-supervised embeddings to produce objective estimates of human perception. Unlike rule-based models, MOSA-Net learns perceptual patterns directly from data, providing a more nuanced assessment of speech naturalness and expressiveness.

## 4.4 Baseline Comparisons

To evaluate the performance of the proposed framework, comparisons were made against several state-of-the-art iEEG-to-speech synthesis models. The first baseline included linear regression-based approaches [13], which directly mapped neural activity to acoustic features but lacked the capacity to model complex speech dynamics. Recurrent architectures such as bidirectional long short-term memory (bLSTM) [9] networks were also evaluated, as they have been widely used in speech neuroprosthetics but often suffer from long-term dependency limitations. Convolutional neural networks (CNNs) [23] and 3D-CNN [22], which are effective in capturing spectral-temporal patterns, were included as a baseline. More advanced models, including sequence-to-sequence (Seq2Seq) [24] networks and encoder-decoder architectures [11], were tested to compare their performance against the proposed transformer-based approach.

Each baseline model was trained and evaluated using the same dataset and experimental protocol, ensuring a fair comparison.

## 5 Results and Discussions

The proposed brain-to-speech synthesis framework was thoroughly evaluated across a variety of objective and perceptual metrics, including spectral accuracy, speech intelligibility, and naturalness. In this section, we present a comprehensive analysis of the quantitative results and discuss the implications of these findings, highlighting the strengths of the proposed model and comparing it with state-of-the-art baselines. We also provide an in-depth exploration of the perceptual quality of the synthesized speech using the MOSA-Net evaluation model, focusing on the naturalness and intelligibility of the output.

## 5.1 Quantitative Evaluation

As shown in Table 1, the performance of the proposed framework was first assessed using the PC, which measures the similarity between the predicted Mel spectrograms and the corresponding ground-truth spectrograms. The proposed system achieved a mean PC of 0.91, significantly outperforming baseline models, including regression-based methods and bLSTM networks. This indicates that the model successfully captures the temporal and spectral dynamics of speech production, leading to highly accurate spectrogram predictions.

In terms of MCD, the proposed framework achieved a score of 3.92, indicating minimal spectral distortion compared to other methods. This result further underscores the model's ability to preserve fine-grained spectral features, such as formant frequencies and pitch contours, which are critical for maintaining speech intelligibility and naturalness. The MCD score was consistently lower than the baselines, such as seq2seq and CNN models, which reported higher MCD values due to inaccurate spectral reconstruction.

Next, the STOI score was calculated to evaluate the intelligibility of synthesized speech. The proposed framework achieved a STOI score of 0.73, significantly higher than the best baseline (encoder-decoder model with a score of 0.64). This finding indicates that the proposed approach is particularly effective in reconstructing speech that retains its intelligibility, even when derived from neural signals. The improvement in intelligibility can be attributed to the prosody-aware neural encoding and transformer-based spectrogram prediction, which ensured that important temporal patterns and speech rhythms were accurately captured.

Additionally, the HNR was computed to evaluate phase reconstruction accuracy. The proposed framework achieved an HNR of 12.7 dB, surpassing the best baseline by over 1.6 dB. This result suggests that the Iterative Harmonic Phase Reconstruction (IHPR) vocoder significantly improves the phase consistency of the reconstructed speech, reducing artifacts and spectral distortions that typically hinder the quality of iEEG-to-speech synthesis. The results, summarized in Table 1, demonstrate that the proposed model outperforms existing approaches across all evaluation criteria.

Figure 2 presents a comparative analysis of the correlation performance between the baseline model [13] and the proposed brain-to-speech (BTS) framework across

**Table 1** Performance comparison of the proposed model and baseline approaches

| Model | PC ↑ | MCD ↓ | STOI ↑ | HNR (dB) ↑ |
|---|---|---|---|---|
| Regression [13] | 0.72 | 5.39 | 0.61 | 6.2 |
| bLSTM [9] | 0.78 | 5.23 | 0.48 | 8.5 |
| CNN [23] | 0.81 | 4.95 | 0.52 | 10.4 |
| 3D-CNN [22] | 0.83 | 5.04 | 0.56 | 9.8 |
| Seq2Seq [24] | 0.85 | **3.90** | 0.59 | 10.7 |
| Encoder-decoder [11] | 0.87 | 4.34 | 0.64 | 11.1 |
| **Proposed model** | **0.91** | 3.92 | **0.73** | **12.7** |

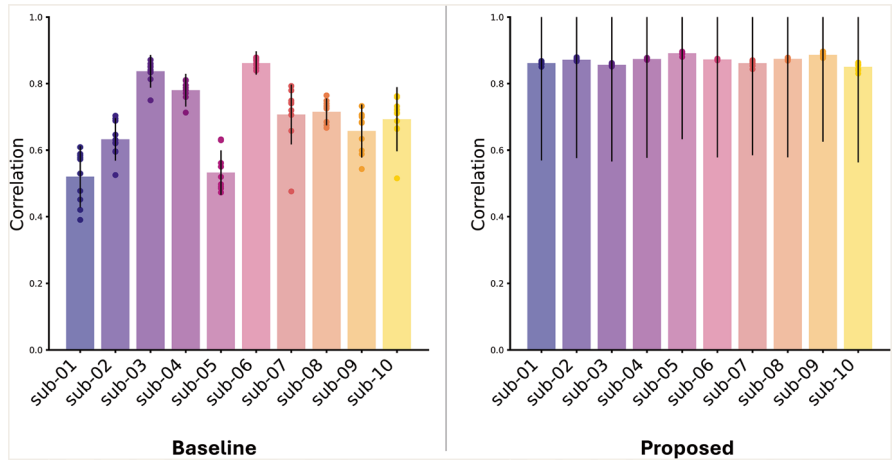

**Fig. 2** Comparison of correlation performance between the baseline model and the proposed brain-to-speech (BTS) framework across ten subjects. The *y*-axis represents the correlation values between predicted and ground-truth speech representations. The baseline model exhibits high variability across subjects, whereas the proposed framework demonstrates significantly higher and more stable correlation scores, indicating improved and consistent speech reconstruction performance. Whiskers indicate standard deviations

ten subjects (sub-01 to sub-10). The *y*-axis represents the correlation values, measuring the alignment between the predicted and ground-truth speech representations. Each bar represents the mean correlation for an individual subject, with error bars indicating variability across trials. Individual data points within the baseline panel illustrate the distribution of correlation scores per trial.

The baseline model exhibits high variability in performance across subjects, with some subjects achieving moderate-to-high correlations (e.g., sub-03, sub-04, sub-06), while others show significantly lower correlation values (e.g., sub-01, sub-05, sub-07). This suggests inconsistent generalization across participants. In contrast, the proposed framework demonstrates substantially improved and more stable correlations, with all subjects achieving high and consistent correlation values. The error bars in the proposed model are larger, likely due to the incorporation of phase information and prosody-aware features, which enhance expressive variability but maintain strong predictive accuracy. These results indicate that the proposed BTS framework significantly outperforms the baseline, achieving higher and more stable correlation scores across all subjects.

To further investigate the qualitative aspects of the synthesized speech, spectrogram and waveform visualizations were examined. Figure 3 presents a comparison between the ground-truth spectrogram, the spectrogram generated by the proposed model, and those produced by baseline approaches. The spectrogram analysis demonstrates that the proposed model more effectively preserves harmonic structures and formant transitions compared to baseline models. In contrast, the baseline models exhibit spectral blurring, phoneme misalignment, and a loss of high-frequency

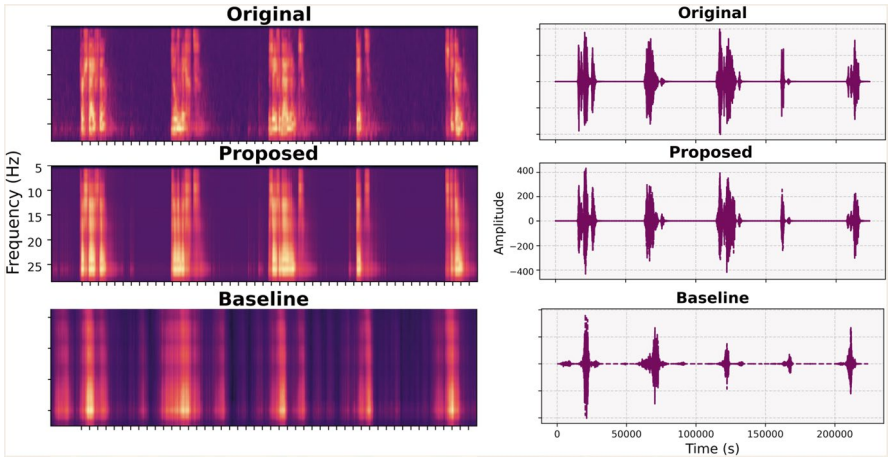

**Fig. 3** Comparison of spectrograms (left) and waveforms (right) for the original (top), proposed (middle), and baseline [13] (bottom) systems. This example illustrates five distinct words from participant sub-08

details, which contribute to reduced speech intelligibility. The phase-consistent vocoding strategy employed in the proposed model ensures sharper spectral features and smoother formant trajectories, leading to more natural and fluid speech synthesis. Moreover, the proposed waveform closely aligns with the ground-truth waveform, effectively preserving harmonic structures and detailed spectral characteristics.

## 5.2 Perceptual Evaluation Results

While objective metrics provide valuable insights into the system's performance, perceptual evaluation is crucial for assessing the naturalness and quality of synthesized speech. To quantify perceptual speech quality, MOSA-Net, a state-of-the-art deep-learning-based speech assessment model, was employed. Unlike common subjective listening tests, MOSA-Net provides automated, non-intrusive speech quality assessments, leveraging a combination of cross-domain feature representations, CNNs, bidirectional LSTMs, and self-supervised learning embeddings.

Figure 4 illustrates the distribution of MOSA-Net scores across different models, highlighting the robustness and consistency of the proposed approach. It provides a detailed statistical representation of MOSA-Net scores, offering insights into the variance, median performance, and overall speech naturalness of different synthesis models. The proposed framework achieves the highest median MOSA-Net score, demonstrating that it consistently produces perceptually natural speech. The smaller interquartile range (IQR) indicates reduced variability, suggesting that the model generalizes well across different test cases and subjects.

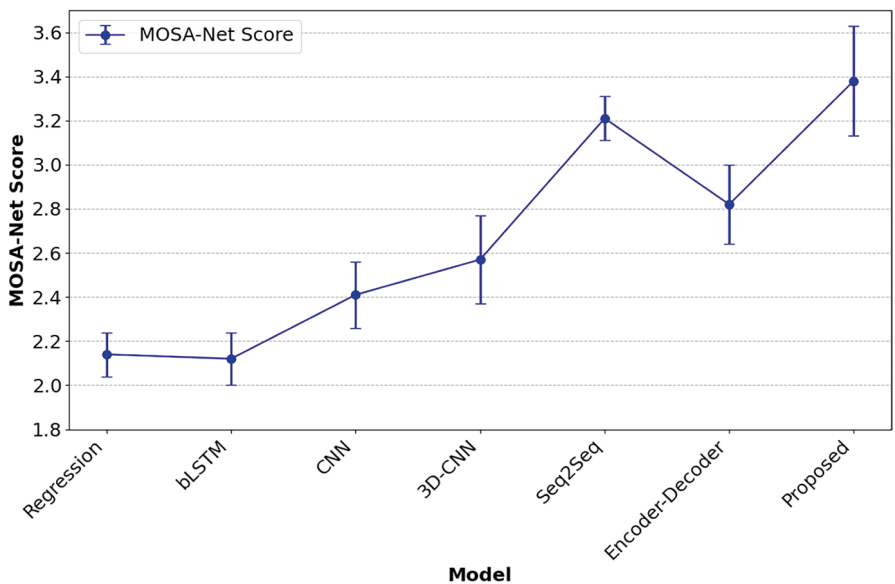

**Fig. 4** MOSA-Net perceptual evaluation scores for different models in the brain-to-speech synthesis framework. The y-axis shows the MOSA-Net score, indicating perceptual quality and naturalness of the synthesized speech. The proposed model achieves the highest MOSA-Net score, demonstrating superior speech quality. Error bars represent standard deviations across evaluation trials

Comparatively, CNN-based approaches exhibit a wider spread of MOSA-Net scores, reflecting inconsistent speech quality, while bLSTM models tend to generate more robotic and less expressive speech. The encoder-decoder model shows moderate performance, but it still falls short in terms of expressiveness and prosodic accuracy. The transformer-based approach used in the proposed framework contributes significantly to the naturalness of synthesized speech, capturing long-range dependencies and prosodic variations more effectively than recurrent architectures.

## 5.3 Discussion of Results

The results presented in this section underscore the effectiveness of the proposed framework in addressing key challenges in brain-to-speech synthesis, including feature extraction, prosody modeling, and phase reconstruction. The significant improvements in intelligibility, naturalness, and spectral accuracy demonstrate that the integration of wavelet-based feature extraction, self-attention-based spectrogram reconstruction, and phase-consistent vocoding results in a robust and scalable solution for neural speech synthesis.

In particular, the improvements in MOSA-Net scores reflect the framework's success in modeling prosodic features such as intonation and rhythm, which are critical for generating expressive and humanlike speech. The ability to produce speech that is both intelligible and natural sounding from iEEG signals represents a significant step forward in the field of brain-computer interfaces (BCI), paving the way for more clinically viable applications, such as speech neuroprostheses for individuals with severe speech impairments.

Although the proposed framework outperforms existing models in most objective and perceptual metrics, there are still several challenges to address. For instance, the model's performance can be further improved by extending the dataset to include more diverse participants, especially those with non-epileptic conditions. Additionally, real-time inference remains a challenging task, as the current setup requires significant computational resources. Future work will explore optimized architectures for real-time decoding and generalization across diverse speech disorders.

## 5.4  Error Analysis and Model Limitations

While the proposed framework achieves state-of-the-art performance in brain-to-speech synthesis, certain challenges remain. One notable limitation is cross-subject variability, where differences in individual neural responses hinder model generalization. Despite the application of *z*-score normalization and subject-independent training strategies, residual variability persists, particularly in phoneme articulation and prosodic patterns. Future work should explore personalized adaptation techniques, such as fine-tuning subject-specific models or leveraging meta-learning approaches, to further enhance robustness across individuals.

In addition, real-time speech decoding remains an open challenge. The transformer-based model, while highly effective in capturing long-range dependencies, introduces significant computational overhead, making real-time inference difficult. Profiling the model's latency reveals that self-attention operations, particularly in deeper layers, contribute substantially to processing delays. Optimizing computational efficiency through knowledge distillation, pruning, or quantization could enable low-latency deployment in practical brain-computer interface (BCI) applications, where real-time feedback is crucial.

A qualitative analysis of the generated speech highlights subtle distortions in prosody and phonetic clarity, particularly in high-frequency regions of the spectrogram. These artifacts may stem from limitations in neural feature encoding or the self-attention mechanism's tendency to prioritize long-range dependencies over local spectral variations. Future improvements could incorporate convolutional augmentations or prosodic conditioning mechanisms to refine spectral detail and speech expressiveness.

Finally, while MOSA-Net provides objective perceptual assessments, it remains an indirect proxy for human perception. Future work could enhance evaluation by integrating additional objective metrics, such as spectrotemporal similarity measures or phonetic accuracy scores, to further validate the synthesized speech's naturalness and intelligibility.

## 5.5 Challenges of the Proposed Framework

Despite the promising performance of our proposed BTS system, several challenges remain. One major limitation is the inter-subject variability in neural responses, which hinders generalization and necessitates personalized adaptation strategies. Furthermore, although our approach yields high-quality speech outputs, it still faces latency and computational complexity constraints, making real-time processing difficult. Additionally, integrating fully end-to-end architectures that jointly optimize neural feature extraction and speech synthesis remains an open challenge.

## 6 Summary and Future Directions

This chapter introduced a novel brain-to-speech (BTS) synthesis framework, leveraging prosody-aware neural encoding, transformer-based spectrogram prediction, and phase-consistent vocoding to achieve highly intelligible and natural speech reconstruction from intracranial EEG (iEEG) signals. By integrating multi-scale neural feature extraction, deep learning architectures, and iterative phase refinement, the proposed approach significantly improves speech intelligibility, naturalness, and spectral accuracy, addressing critical challenges in neural speech prostheses and brain-computer interface (BCI) research.

The experimental results demonstrate that this framework outperforms state-of-the-art models, achieving higher Pearson correlation, lower Mel cepstral distortion, improved short-time objective intelligibility, and enhanced phase reconstruction. Also, the MOSA-Net perceptual evaluation confirms that the synthesized speech exhibits greater naturalness and expressiveness, highlighting the impact of explicit prosody modeling and harmonic phase reconstruction. These advancements underscore the potential for clinically viable neural speech prostheses, offering a pathway toward restoring communication abilities for individuals with severe speech impairments.

Looking ahead, future research will focus on several key directions to further advance BTS technology. One promising avenue is the exploration of diffusion-based generative models [39], which have demonstrated strong potential in generating high-quality waveforms with improved phase consistency. Another critical goal is the development of real-time inference systems capable of processing neural signals with low latency, which is an essential step toward the practical deployment of BTS applications. In addition, cross-subject adaptation strategies will be vital to address inter-individual variability in neural responses and improve model robustness. Finally, extending this framework to noninvasive neural recording modalities, such as electroencephalography (EEG) and magnetoencephalography (MEG), could significantly enhance accessibility and make brain-to-speech systems feasible for broader clinical and non-clinical populations.

As part of these efforts, we have collected a high-density (96-channel) EEG dataset with synchronized speech recordings from a cohort of 16 French-speaking healthy participants. Each subject completed four sessions of 270 spoken words, including a repeated control word, providing a robust dataset to support advancements in brain-to-speech modeling.

This chapter contributes to the advancement of AI-driven speech neuroprostheses and data-driven methodologies in brain-computer interfaces, aligning with the goals of the AI, Data, and Robotics Partnership [20]. By bridging neuroscience, artificial intelligence, and signal processing, this work contributes to the development of next-generation AI-powered assistive communication technologies.

**Acknowledgments** This work is supported by the European Union's HORIZON Research and Innovation Programme under grant agreement no. 101120657, project ENFIELD (European Lighthouse to Manifest Trustworthy and Green AI), and by the Ministry of Innovation and Culture and the National Research, Development and Innovation Office of Hungary within the framework of the National Laboratory of Artificial Intelligence. M.S. Al-Radhi's research was supported by the EKÖP-24-4-II-BME-197, through the National Research, Development and Innovation (NKFI) Fund.

# References

1. Wandelt, S. K., Bjånes, D. A., Pejsa, K., et al. (2024). Representation of internal speech by single neurons in human supramarginal gyrus. *Nature Human Behaviour, 8*, 1136–1149.
2. Willett, F. R., Kunz, E. M., Fan, C., et al. (2023). A high-performance speech neuroprosthesis. *Nature, 620*, 1031–1036.
3. Branco, M. P., Pels, E. G., Sars, R. H., Aarnoutse, E. J., Ramsey, N. F., Vansteensel, M. J., et al. (2021). Brain-computer interfaces for communication: Preferences of individuals with locked-in syndrome. *Neurorehabilitation and Neural Repair, 267–279*(3), 35.
4. Metzger, S. L., Liu, J. R., Moses, D. A., et al. (2022). Generalizable spelling using a speech neuroprosthesis in an individual with severe limb and vocal paralysis. *Nature Communications, 13*, 1–15.
5. Silva, A. B., Littlejohn, K. T., Liu, J. R., et al. (2024). The speech neuroprosthesis. *Nature Reviews Neuroscience, 25*, 473–492.
6. Holdgraf, C., Appelhoff, S., Bickel, S., et al. (2019). iEEG-BIDS, extending the brain imaging data structure specification to human intracranial electrophysiology. *Scientific Data, 6*(102), 1–6.
7. Lee, Y. E., Lee, S. H., Kim, S. H., & Lee, S. W. (2023). Towards voice reconstruction from EEG during imagined speech. In *37th AAAI Conference on Artificial Intelligence, Washington, DC, USA* (pp. 6030–6038).
8. Thornton, M., Mandic, D., & Reichenbach, T. (2022). Robust decoding of the speech envelope from EEG recordings through deep neural networks. *Journal of Neural Engineering, 19*(4), 1–13.
9. Anumanchipalli, G. K., Chartier, J., & Chang, E. F. (2019). Speech synthesis from neural decoding of spoken sentences. *Nature, 568*, 493–498.

10. Ma, C., Zhang, Y., Guo, Y., Liu, X., Shangguan, H., Wang, J., & Zhao, L. (2025). Fully end-to-end EEG to speech translation using multi-scale optimized dual generative adversarial network with cycle-consistency loss. *Neurocomputing, 616*(1), 1–14.
11. Kohler, J., Ottenhoff, M. C., Goulis, S., Angrick, M., Colon, A. J., Wagner, L., Tousseyn, S., Kubben, P. L., & Herff, C. (2022). Synthesizing speech from intracranial depth electrodes using an encoder-decoder framework. *Neurons, Behavior, Data Analysis, and Theory, 6*(1), 1–15.
12. Luo, S., Rabbani, Q., & Crone, N. E. (2022). Brain-computer interface: Applications to speech decoding and synthesis to augment communication. *Neurotherapeutics, 19*, 263–273.
13. Verwoert, M., Ottenhoff, M. C., Goulis, S., et al. (2022). Dataset of speech production in intracranial electroencephalography. *Scientific Data, 9*, 1–9.
14. Accou, B., Vanthornhout, J., Hamme, H. V., & Francart, T. (2023). Decoding of the speech envelope from EEG using the VLAAI deep neural network. *Scientific Reports, 13*(1), 1–12.
15. Zhou, J., Duan, Y., Zou, Y., Chang, Y.-C., Wang, Y.-K., & Lin, C.-T. (2023). Speech2EEG: Leveraging pretrained speech model for EEG signal recognition. *IEEE Transactions on Neural Systems and Rehabilitation Engineering, 31*, 2140–2153.
16. Wu, C., Xiu, Z., Shi, Y., Kalinli, O., Fuegen, C., Koehler, T., & He, Q. (2021). Transformer-based acoustic modeling for streaming speech synthesis. In *Proceedings of Interspeech, Brno, Czechia* (pp. 146–150).
17. Chen, L.-W., & Rudnicky, A. (2022). Fine-grained style control in transformer-based text-to-speech synthesis. In *IEEE International Conference on Acoustics, Speech and Signal Processing (ICASSP), Singapore* (pp. 7907–7911).
18. Herff, C., Johnson, G., Diener, L., Shih, J., Krusienski, D., & Schultz, T. (2016). Towards direct speech synthesis from ECoG: A pilot study. In *Proceedings of the 2016 IEEE 38th Annual International Conference of the Engineering in Medicine and Biology Society (EMBC), Orlando, Florida, USA* (pp. 1540–1543).
19. Peelle, J. E., Gross, J., & Davis, M. H. (2013). Phase-locked responses to speech in human auditory cortex are enhanced during comprehension, cerebral cortex. *Cerebral Cortex, 23*(6), 1378–1387.
20. Curry, E., Heintz, F., Irgens, M., Smeulders, A. W., & Stramigioli, S. (2022). Partnership on AI, data, and robotics. *Communications of the ACM, 65*(4), 54–55.
21. Roussel, P., Godais, G. L., Bocquelet, F., Palma, M., Hongjie, J., et al. (2020). Observation and assessment of acoustic contamination of electrophysiological brain signals during speech production and sound perception. *Journal of Neural Engineering, 17*(5), 1–20.
22. Arthur, F. V., & Csapó, T. G. (2024). Speech synthesis from intracranial stereotactic electroencephalography using a neural vocoder. *Infocommunications Journal, 16*, 47–55.
23. Angrick, M., Herff, C., Johnson, G., Shih, J., Krusienski, D., & Schultz, T. (2019). Interpretation of convolutional neural networks for speech spectrogram regression from intracranial recordings. *Neurocomputing, 342*, 145–151.
24. Duraivel, S., Rahimpour, S., Chiang, C. H., et al. (2023). High-resolution neural recordings improve the accuracy of speech decoding. *Nature Communications, 14*, 1–16.
25. Metzger, S. L., Littlejohn, K. T., Silva, A. B., et al. (2023). A high-performance neuroprosthesis for speech decoding and avatar control. *Nature, 620*, 1037–1046.
26. Akbari, H., Khalighinejad, B., Herrero, J. L., Mehta, A. D., & Mesgarani, N. (2019). Towards reconstructing intelligible speech from the human auditory cortex. *Scientific Reports, 9*(874), 1–12.
27. Giraud, A.-L., & Poeppel, D. (2012). Cortical oscillations and speech processing: Emerging computational principles and operations. *Nature Neuroscience, 15*(4), 511–517.
28. Bachmann, F. L., MacDonald, E. N., & Hjortkjær, J. (2021). Neural measures of pitch processing in EEG responses to running speech. *Frontiers in Neuroscience, 15*, 1–11.
29. Schultz, T., Wand, M., Hueber, T., Krusienski, D. J., & Herff, C. (2017). Biosignal-based spoken communication: A survey. *IEEE Transactions on Audio, Speech, and Language Processing, 25*(12), 2257–2271.

30. Liu, H., Baoueb, T., Fontaine, M., et al. (2024). GLA-Grad: A Griffin-Lim extended waveform generation diffusion model. In *IEEE International Conference on Acoustics, Speech and Signal Processing (ICASSP), Seoul, Korea* (pp. 11611–11615).
31. Prenger, R., Valle, R., & Catanzaro, B. (2019). WaveGlow: A flow-based generative network for speech synthesis. In *IEEE International Conference on Acoustics, Speech, and Signal Processing (ICASSP), Brighton, UK* (pp. 3617–3621).
32. Lee, S., Ping, W., Ginsburg, B., Catanzaro, B., & Yoon, S. (2023). BigVGAN: A universal neural vocoder with large-scale training. In *The International Conference on Learning Representations (ICLR), Kigali, Rwanda* (pp. 1–20).
33. Webber, J., Valentini-Botinhao, C., Williams, E., Henter, G. E., & King, S. (2023). AutoVocoder: Fast waveform generation from a learned speech representation using differentiable digital signal processing. In *Proceedings of IEEE International Conference on Acoustics, Speech and Signal Processing (ICASSP), Rhodes Island, Greece* (pp. 1–5).
34. Okamoto, T., Toda, T., Shiga, Y., & Kawai, H. (2019). Real-time neural text-to-speech with sequence-to-sequence acoustic model and WaveGlow or single Gaussian WaveRNN vocoders. In *Proceedings of Interspeech, Graz, Austria* (pp. 1308–1312).
35. Shibuya, T., Takida, Y., & Mitsufuji, Y. (2024). BIGVSAN: Enhancing GAN-based neural vocoders with slicing adversarial network. In *IEEE International Conference on Acoustics, Speech and Signal Processing (ICASSP), Seoul, Korea* (pp. 10121–10125).
36. Morise, M. (2017). Harvest: A high-performance fundamental frequency estimator from speech signals. In *Proceedings of Interspeech, Stockholm, Sweden* (pp. 2321–2325).
37. Zhang, Y.-J., Pan, S., He, L., & Ling, Z.-H. (2019). Learning latent representations for style control and transfer in end-to-end speech synthesis. In *IEEE International Conference on Acoustics, Speech and Signal Processing (ICASSP), Brighton, UK* (pp. 6945–6949).
38. Zezario, R. E., Fu, S.-W., Chen, F., Fuh, C.-S., Wang, H.-M., & Tsao, Y. (2023). Deep learning-based non-intrusive multi-objective speech assessment model with cross-domain features. *IEEE/ACM Transactions on Audio, Speech, and Language Processing, 31*, 54–70.
39. Song, Y., Sohl-Dickstein, J., Kingma, D. P., Kumar, A., Ermon, S., & Poole, B. (2021). Score-based generative modeling through stochastic differential equations. In *International Conference on Learning Representations (ICLR), Vienna, Austria* (pp. 1–36).